\documentstyle[preprint,aps,prl]{revtex}

\def\temp{1.35}%
\let\tempp=\relax
\expandafter\ifx\csname psboxversion\endcsname\relax
\else
    \ifdim\temp cm>\psboxversion cm
    \else
      \let\temp=\psboxversion
      \let\tempp= 
    \fi
\fi
\tempp
\let\psboxversion=\temp
\catcode`\@=11
\def\psfortextures{
\def\PSspeci@l##1##2{%
\special{illustration ##1\space scaled ##2}%
}}%
\def\psfordvitops{
\def\PSspeci@l##1##2{%
\special{dvitops: import ##1\space \the\drawingwd \the\drawinght}%
}}%
\def\psfordvips{
\def\PSspeci@l##1##2{%
\d@my=0.1bp \d@mx=\drawingwd \divide\d@mx by\d@my
\includegraphics{##1\space}}}%
\def\psforoztex{
\def\PSspeci@l##1##2{%
\special{##1 \space
      ##2 1000 div dup scale
      \number-\psllx\space\space \number-\pslly\space\space translate
}}}%
\def\psfordvitps{
\def\dvitpsLiter@ldim##1{\dimen0=##1\relax
\special{dvitps: Literal "\number\dimen0\space"}}%
\def\PSspeci@l##1##2{%
\at(0bp;\drawinght){%
\special{dvitps: Include0 "psfig.psr"}
\dvitpsLiter@ldim{\drawingwd}%
\dvitpsLiter@ldim{\drawinght}%
\dvitpsLiter@ldim{\psllx bp}%
\dvitpsLiter@ldim{\pslly bp}%
\dvitpsLiter@ldim{\psurx bp}%
\dvitpsLiter@ldim{\psury bp}%
\special{dvitps: Literal "startTexFig"}%
\special{dvitps: Include1 "##1"}%
\special{dvitps: Literal "endTexFig"}%
}}}%
\def\psfordvialw{
\def\PSspeci@l##1##2{
\special{language "PostScript",
position = "bottom left",
literal "  \psllx\space \pslly\space translate
  ##2 1000 div dup scale
  -\psllx\space -\pslly\space translate",
include "##1"}
}}%
\def\psforptips{
\def\PSspeci@l##1##2{{
\d@mx=\psurx bp
\advance \d@mx by -\psllx bp
\divide \d@mx by 1000\multiply\d@mx by \xscale
\incm{\d@mx}
\let\tmpx\dimincm
\d@my=\psury bp
\advance \d@my by -\pslly bp
\divide \d@my by 1000\multiply\d@my by \xscale
\incm{\d@my}
\let\tmpy\dimincm
\d@mx=-\psllx bp
\divide \d@mx by 1000\multiply\d@mx by \xscale
\d@my=-\pslly bp
\divide \d@my by 1000\multiply\d@my by \xscale
\at(\d@mx;\d@my){\special{ps:##1 x=\tmpx cm, y=\tmpy cm}}
}}}%
\def\psonlyboxes{
\def\PSspeci@l##1##2{%
\at(0cm;0cm){\boxit{\vbox to\drawinght
  {\vss\hbox to\drawingwd{\at(0cm;0cm){\hbox{({\tt##1})}}\hss}}}}
}}%
\def\psloc@lerr#1{%
\let\savedPSspeci@l=\PSspeci@l%
\def\PSspeci@l##1##2{%
\at(0cm;0cm){\boxit{\vbox to\drawinght
  {\vss\hbox to\drawingwd{\at(0cm;0cm){\hbox{({\tt##1}) #1}}\hss}}}}
\let\PSspeci@l=\savedPSspeci@l
}}%
\newread\pst@mpin
\newdimen\drawinght\newdimen\drawingwd
\newdimen\psxoffset\newdimen\psyoffset
\newbox\drawingBox
\newcount\xscale \newcount\yscale \newdimen\pscm\pscm=1cm
\newdimen\d@mx \newdimen\d@my
\newdimen\pswdincr \newdimen\pshtincr
\let\ps@nnotation=\relax
{\catcode`\|=0 |catcode`|\=12 |catcode`|
|catcode`#=12 |catcode`*=14
|xdef|backslashother{\}*
|xdef|percentother{
|xdef|tildeother{~}*
|xdef|sharpother{#}*
}%
\def\R@moveMeaningHeader#1:->{}%
\def\uncatcode#1{%
\edef#1{\expandafter\R@moveMeaningHeader\meaning#1}}%
\def\execute#1{#1}
\def\psm@keother#1{\catcode`#112\relax}
\def\executeinspecs#1{%
\execute{\begingroup\let\do\psm@keother\dospecials\catcode`\^^M=9#1\endgroup}}%
\def\@mpty{}%
\def\matchexpin#1#2{
  \fi%
  \edef\tmpb{{#2}}%
  \expandafter\makem@tchtmp\tmpb%
  \edef\tmpa{#1}\edef\tmpb{#2}%
  \expandafter\expandafter\expandafter\m@tchtmp\expandafter\tmpa\tmpb\endm@tch%
  \if\match%
}%
\def\matchin#1#2{%
  \fi%
  \makem@tchtmp{#2}%
  \m@tchtmp#1#2\endm@tch%
  \if\match%
}%
\def\makem@tchtmp#1{\def\m@tchtmp##1#1##2\endm@tch{%
  \def\tmpa{##1}\def\tmpb{##2}\let\m@tchtmp=\relax%
  \ifx\tmpb\@mpty\def\match{YN}%
  \else\def\match{YY}\fi%
}}%
\def\incm#1{{\psxoffset=1cm\d@my=#1
 \d@mx=\d@my
  \divide\d@mx by \psxoffset
  \xdef\dimincm{\number\d@mx.}
  \advance\d@my by -\number\d@mx cm
  \multiply\d@my by 100
 \d@mx=\d@my
  \divide\d@mx by \psxoffset
  \edef\dimincm{\dimincm\number\d@mx}
  \advance\d@my by -\number\d@mx cm
  \multiply\d@my by 100
 \d@mx=\d@my
  \divide\d@mx by \psxoffset
  \xdef\dimincm{\dimincm\number\d@mx}
}}%
\newif\ifNotB@undingBox
\newhelp\PShelp{Proceed: you'll have a 5cm square blank box instead of
your graphics.}%
\def\s@tsize#1 #2 #3 #4\@ndsize{
  \def\psllx{#1}\def\pslly{#2}%
  \def\psurx{#3}\def\psury{#4}
  \ifx\psurx\@mpty\NotB@undingBoxtrue
  \else
    \drawinght=#4bp\advance\drawinght by-#2bp
    \drawingwd=#3bp\advance\drawingwd by-#1bp
  \fi
  }%
\def\sc@nBBline#1:#2\@ndBBline{\edef\p@rameter{#1}\edef\v@lue{#2}}%
\def\g@bblefirstblank#1#2:{\ifx#1 \else#1\fi#2}%
{\catcode`\%=12
\xdef\B@undingBox{
\def\ReadPSize#1{
 \readfilename#1\relax
 \let\PSfilename=\lastreadfilename
 \openin\pst@mpin=#1\relax
 \ifeof\pst@mpin \errhelp=\PShelp
   \errmessage{I haven't found your postscript file (\PSfilename)}%
   \psloc@lerr{was not found}%
   \s@tsize 0 0 142 142\@ndsize
   \closein\pst@mpin
 \else
   \if\matchexpin{\GlobalInputList}{, \lastreadfilename}%
   \else\xdef\GlobalInputList{\GlobalInputList, \lastreadfilename}%
     \immediate\write\psbj@inaux{\lastreadfilename,}%
   \fi%
   \loop
     \executeinspecs{\catcode`\ =10\global\read\pst@mpin to\n@xtline}%
     \ifeof\pst@mpin
       \errhelp=\PShelp
       \errmessage{(\PSfilename) is not an Encapsulated PostScript File:
           I could not find any \B@undingBox: line.}%
       \edef\v@lue{0 0 142 142:}%
       \psloc@lerr{is not an EPSFile}%
       \NotB@undingBoxfalse
     \else
       \expandafter\sc@nBBline\n@xtline:\@ndBBline
       \ifx\p@rameter\B@undingBox\NotB@undingBoxfalse
         \edef\t@mp{%
           \expandafter\g@bblefirstblank\v@lue\space\space\space}%
         \expandafter\s@tsize\t@mp\@ndsize
       \else\NotB@undingBoxtrue
       \fi
     \fi
   \ifNotB@undingBox\repeat
   \closein\pst@mpin
 \fi
\message{#1}%
}%
\def\psboxto(#1;#2)#3{\vbox{%
   \ReadPSize{#3}%
   \advance\pswdincr by \drawingwd
   \advance\pshtincr by \drawinght
   \divide\pswdincr by 1000
   \divide\pshtincr by 1000
   \d@mx=#1
   \ifdim\d@mx=0pt\xscale=1000
         \else \xscale=\d@mx \divide \xscale by \pswdincr\fi
   \d@my=#2
   \ifdim\d@my=0pt\yscale=1000
         \else \yscale=\d@my \divide \yscale by \pshtincr\fi
   \ifnum\yscale=1000
         \else\ifnum\xscale=1000\xscale=\yscale
                    \else\ifnum\yscale<\xscale\xscale=\yscale\fi
              \fi
   \fi
   \divide\drawingwd by1000 \multiply\drawingwd by\xscale
   \divide\drawinght by1000 \multiply\drawinght by\xscale
   \divide\psxoffset by1000 \multiply\psxoffset by\xscale
   \divide\psyoffset by1000 \multiply\psyoffset by\xscale
   \global\divide\pscm by 1000
   \global\multiply\pscm by\xscale
   \multiply\pswdincr by\xscale \multiply\pshtincr by\xscale
   \ifdim\d@mx=0pt\d@mx=\pswdincr\fi
   \ifdim\d@my=0pt\d@my=\pshtincr\fi
   \message{scaled \the\xscale}%
 \hbox to\d@mx{\hss\vbox to\d@my{\vss
   \global\setbox\drawingBox=\hbox to 0pt{\kern\psxoffset\vbox to 0pt{%
      \kern-\psyoffset
      \PSspeci@l{\PSfilename}{\the\xscale}%
      \vss}\hss\ps@nnotation}%
   \global\wd\drawingBox=\the\pswdincr
   \global\ht\drawingBox=\the\pshtincr
   \global\drawingwd=\pswdincr
   \global\drawinght=\pshtincr
   \baselineskip=0pt
   \copy\drawingBox
 \vss}\hss}%
  \global\psxoffset=0pt
  \global\psyoffset=0pt
  \global\pswdincr=0pt
  \global\pshtincr=0pt 
  \global\pscm=1cm 
}}%
\def\psboxscaled#1#2{\vbox{%
  \ReadPSize{#2}%
  \xscale=#1
  \message{scaled \the\xscale}%
  \divide\pswdincr by 1000 \multiply\pswdincr by \xscale
  \divide\pshtincr by 1000 \multiply\pshtincr by \xscale
  \divide\psxoffset by1000 \multiply\psxoffset by\xscale
  \divide\psyoffset by1000 \multiply\psyoffset by\xscale
  \divide\drawingwd by1000 \multiply\drawingwd by\xscale
  \divide\drawinght by1000 \multiply\drawinght by\xscale
  \global\divide\pscm by 1000
  \global\multiply\pscm by\xscale
  \global\setbox\drawingBox=\hbox to 0pt{\kern\psxoffset\vbox to 0pt{%
     \kern-\psyoffset
     \PSspeci@l{\PSfilename}{\the\xscale}%
     \vss}\hss\ps@nnotation}%
  \advance\pswdincr by \drawingwd
  \advance\pshtincr by \drawinght
  \global\wd\drawingBox=\the\pswdincr
  \global\ht\drawingBox=\the\pshtincr
  \global\drawingwd=\pswdincr
  \global\drawinght=\pshtincr
  \baselineskip=0pt
  \copy\drawingBox
  \global\psxoffset=0pt
  \global\psyoffset=0pt
  \global\pswdincr=0pt
  \global\pshtincr=0pt 
  \global\pscm=1cm
}}%
\def\psbox#1{\psboxscaled{1000}{#1}}%
\newif\ifn@teof\n@teoftrue
\newif\ifc@ntrolline
\newif\ifmatch
\newread\j@insplitin
\newwrite\j@insplitout
\newwrite\psbj@inaux
\immediate\openout\psbj@inaux=psbjoin.aux
\immediate\write\psbj@inaux{\string\joinfiles}%
\immediate\write\psbj@inaux{\jobname,}%
\def\toother#1{\ifcat\relax#1\else\expandafter%
  \toother@ux\meaning#1\endtoother@ux\fi}%
\def\toother@ux#1 #2#3\endtoother@ux{\def\tmp{#3}%
  \ifx\tmp\@mpty\def\tmp{#2}\let\next=\relax%
  \else\def\next{\toother@ux#2#3\endtoother@ux}\fi%
\next}%
\let\readfilenamehook=\relax
\def\re@d{\expandafter\re@daux}
\def\re@daux{\futurelet\nextchar\stopre@dtest}%
\def\re@dnext{\xdef\lastreadfilename{\lastreadfilename\nextchar}%
  \afterassignment\re@d\let\nextchar}%
\def\stopre@d{\egroup\readfilenamehook}%
\def\stopre@dtest{%
  \ifcat\nextchar\relax\let\nextread\stopre@d
  \else
    \ifcat\nextchar\space\def\nextread{%
      \afterassignment\stopre@d\chardef\nextchar=`}%
    \else\let\nextread=\re@dnext
      \toother\nextchar
      \edef\nextchar{\tmp}%
    \fi
  \fi\nextread}%
\def\readfilename{\bgroup%
  \let\\=\backslashother \let\%=\percentother \let\~=\tildeother
  \let\#=\sharpother \xdef\lastreadfilename{}%
  \re@d}%
\xdef\GlobalInputList{\jobname}%
\def\psnewinput{%
  \def\readfilenamehook{
    \if\matchexpin{\GlobalInputList}{, \lastreadfilename}%
    \else\xdef\GlobalInputList{\GlobalInputList, \lastreadfilename}%
      \immediate\write\psbj@inaux{\lastreadfilename,}%
    \fi%
    \let\readfilenamehook=\relax%
    \ps@ldinput\lastreadfilename\relax%
  }\readfilename%
}%
\expandafter\ifx\csname @@input\endcsname\relax    
  \immediate\let\ps@ldinput=\input\def\input{\psnewinput}%
\else
  \immediate\let\ps@ldinput=\@@input
  \def\@@input{\psnewinput}%
\fi%
\def\nowarnopenout{%
 \def\warnopenout##1##2{%
   \readfilename##2\relax
   \message{\lastreadfilename}%
   \immediate\openout##1=\lastreadfilename\relax}}%
\def\warnopenout#1#2{%
 \readfilename#2\relax
 \def\t@mp{TrashMe,psbjoin.aux,psbjoint.tex,}\uncatcode\t@mp
 \if\matchexpin{\t@mp}{\lastreadfilename,}%
 \else
   \immediate\openin\pst@mpin=\lastreadfilename\relax
   \ifeof\pst@mpin
     \else
     \edef\tmp{{If the content of this file is precious to you, this
is your last chance to abort (ie press x or e) and rename it before
retexing (\jobname). If you're sure there's no file
(\lastreadfilename) in the directory of (\jobname), then go on: I'm
simply worried because you have another (\lastreadfilename) in some
directory I'm looking in for inputs...}}%
     \errhelp=\tmp
     \errmessage{I may be about to replace your file named \lastreadfilename}%
   \fi
   \immediate\closein\pst@mpin
 \fi
 \message{\lastreadfilename}%
 \immediate\openout#1=\lastreadfilename\relax}%
{\catcode`\%=12\catcode`\*=14
\gdef\splitfile#1{*
 \readfilename#1\relax
 \immediate\openin\j@insplitin=\lastreadfilename\relax
 \ifeof\j@insplitin
   \message{! I couldn't find and split \lastreadfilename!}*
 \else
   \immediate\openout\j@insplitout=TrashMe
   \message{< Splitting \lastreadfilename\space into}*
   \loop
     \ifeof\j@insplitin
       \immediate\closein\j@insplitin\n@teoffalse
     \else
       \n@teoftrue
       \executeinspecs{\global\read\j@insplitin to\spl@tinline\expandafter
         \ch@ckbeginnewfile\spl@tinline
       \ifc@ntrolline
       \else
         \toks0=\expandafter{\spl@tinline}*
         \immediate\write\j@insplitout{\the\toks0}*
       \fi
     \fi
   \ifn@teof\repeat
   \immediate\closeout\j@insplitout
 \fi\message{>}*
}*
\gdef\ch@ckbeginnewfile#1
 \def\t@mp{#1}*
 \ifx\@mpty\t@mp
   \def\t@mp{#3}*
   \ifx\@mpty\t@mp
     \global\c@ntrollinefalse
   \else
     \immediate\closeout\j@insplitout
     \warnopenout\j@insplitout{#2}*
     \global\c@ntrollinetrue
   \fi
 \else
   \global\c@ntrollinefalse
 \fi}*
\gdef\joinfiles#1\into#2{*
 \message{< Joining following files into}*
 \warnopenout\j@insplitout{#2}*
 \message{:}*
 {*
 \edef\w@##1{\immediate\write\j@insplitout{##1}}*
\w@{
\w@{
\w@{
\w@{
\w@{
\w@{
\w@{
\w@{
\w@{
\w@{
\w@{\string\input\space psbox.tex}*
\w@{\string\splitfile{\string\jobname}}*
\w@{\string\let\string\autojoin=\string\relax}*
}*
 \expandafter\tre@tfilelist#1, \endtre@t
 \immediate\closeout\j@insplitout
 \message{>}*
}*
\gdef\tre@tfilelist#1, #2\endtre@t{*
 \readfilename#1\relax
 \ifx\@mpty\lastreadfilename
 \else
   \immediate\openin\j@insplitin=\lastreadfilename\relax
   \ifeof\j@insplitin
     \errmessage{I couldn't find file \lastreadfilename}*
   \else
     \message{\lastreadfilename}*
     \immediate\write\j@insplitout{
     \executeinspecs{\global\read\j@insplitin to\oldj@ininline}*
     \loop
       \ifeof\j@insplitin\immediate\closein\j@insplitin\n@teoffalse
       \else\n@teoftrue
         \executeinspecs{\global\read\j@insplitin to\j@ininline}*
         \toks0=\expandafter{\oldj@ininline}*
         \let\oldj@ininline=\j@ininline
         \immediate\write\j@insplitout{\the\toks0}*
       \fi
     \ifn@teof
     \repeat
   \immediate\closein\j@insplitin
   \fi
   \tre@tfilelist#2, \endtre@t
 \fi}*
}%
\def\autojoin{%
 \immediate\write\psbj@inaux{\string\into{psbjoint.tex}}%
 \immediate\closeout\psbj@inaux
 \expandafter\joinfiles\GlobalInputList\into{psbjoint.tex}%
}%
\def\centinsert#1{\midinsert\line{\hss#1\hss}\endinsert}%
\def\psannotate#1#2{\vbox{%
  \def\ps@nnotation{#2\global\let\ps@nnotation=\relax}#1}}%
\def\pscaption#1#2{\vbox{%
   \setbox\drawingBox=#1
   \copy\drawingBox
   \vskip\baselineskip
   \vbox{\hsize=\wd\drawingBox\setbox0=\hbox{#2}%
     \ifdim\wd0>\hsize
       \noindent\unhbox0\tolerance=5000
    \else\centerline{\box0}%
    \fi
}}}%
\def\at(#1;#2)#3{\setbox0=\hbox{#3}\ht0=0pt\dp0=0pt
  \rlap{\kern#1\vbox to0pt{\kern-#2\box0\vss}}}%
\newdimen\gridht \newdimen\gridwd
\def\gridfill(#1;#2){%
  \setbox0=\hbox to 1\pscm
  {\vrule height1\pscm width.4pt\leaders\hrule\hfill}%
  \gridht=#1
  \divide\gridht by \ht0
  \multiply\gridht by \ht0
  \gridwd=#2
  \divide\gridwd by \wd0
  \multiply\gridwd by \wd0
  \advance \gridwd by \wd0
  \vbox to \gridht{\leaders\hbox to\gridwd{\leaders\box0\hfill}\vfill}}%
\def\fillinggrid{\at(0cm;0cm){\vbox{%
  \gridfill(\drawinght;\drawingwd)}}}%
\def\textleftof#1:{%
  \setbox1=#1
  \setbox0=\vbox\bgroup
    \advance\hsize by -\wd1 \advance\hsize by -2em}%
\def\textrightof#1:{%
  \setbox0=#1
  \setbox1=\vbox\bgroup
    \advance\hsize by -\wd0 \advance\hsize by -2em}%
\def\endtext{%
  \egroup
  \hbox to \hsize{\valign{\vfil##\vfil\cr%
\box0\cr%
\noalign{\hss}\box1\cr}}}%
\def\frameit#1#2#3{\hbox{\vrule width#1\vbox{%
  \hrule height#1\vskip#2\hbox{\hskip#2\vbox{#3}\hskip#2}%
        \vskip#2\hrule height#1}\vrule width#1}}%
\def\boxit#1{\frameit{0.4pt}{0pt}{#1}}%
\catcode`\@=12 
\psfordvips   

\begin{document}

\draft

\title{Collective excitations of atomic Bose--Einstein condensates}
\author{Mark Edwards\cite{NIST}}
\address{Department of Physics,
Georgia Southern University,
Statesboro, GA 30460-8031.}
\author{P.\ A.\ Ruprecht and K.\ Burnett\cite{NIST}}
\address{Clarendon Laboratory, Department of Physics,
University of Oxford,\\
Parks Road, Oxford OX1 3PU, United Kingdom.}
\author{R.\ J.\ Dodd\cite{IPST} and  Charles\ W.\ Clark}
\address{Electron and Optical Physics Division,
National Institute of Standards and Technology,\\
Technology Administration,
U. S. Department of Commerce,
Gaithersburg, MD 20899.}

\date{\today}

\maketitle

\begin{abstract}
We apply linear--response analysis of the Gross--Pitaevskii 
equation to obtain the excitation frequencies of a 
Bose--Einstein condensate confined in a time--averaged orbiting 
potential trap.  Our calculated values are in excellent agreement 
with those observed in a recent experiment.
\end{abstract}
\draft

\narrowtext

The recent attainment of quantum degeneracy conditions in 
magnetically trapped alkali vapors~\cite{BEC!,Hulet,Ketterle} 
has opened the road to understanding of the many--body physics 
of Bose-Einstein condensates (BECs) in unprecedented detail.  
For dilute gases, it is believed that the essential physics of 
the BEC ground state is captured in the Gross--Pitaevski 
mean--field formalism.  Calculations done with the 
Gross--Pitaevski (GP) equation~\cite{HC96,EDCRB96} have indeed 
agreed reasonably well with the few experimental determinations 
of condensate shapes, sizes, and lifetimes that have been made 
to date, but it cannot be said that the theory has been subject 
to stringent tests.

In this paper we report theoretical results for the excitation
spectrum of a BEC of trapped $^{87}$Rb, obtained by computing 
the response to small mechanical disturbances of a BEC described 
by the GP equation~\cite{liyou}.  These results are compared with 
those of a recent experiment \cite{JILA}, which has observed the free 
oscillations of a BEC that is briefly shaken at frequencies
near resonance.  We believe that this comparison provides the 
most critical quantitative test of mean-field theory made to date.  
The agreement between experimental and theoretical results is 
excellent.  This suggests that the GP equation and its variants 
can adequately describe excited-state as well as ground--state 
properties, and so should provide a practical framework for 
explorations of emerging issues of condensate dynamics, 
finite--temperature phenomena, and atom laser~\cite{atom_laser} 
design.
 
To describe a magnetically trapped atomic gas, we adopt the standard 
GP equation, which is applicable~\cite{griffin,Fetter} when the
condensate fraction of a gaseous system is close to unity.  Each 
atom in the condensate occupies the same orbital $\psi_{g}({\bf r})$, 
which is determined by solution of the nonlinear Schr\"odinger equation,
\begin{equation}
\left[H_{0} + N_{0}U_{0}|\psi_{g}({\bf r})|^{2}\right]\psi_{g}({\bf r}) = 
\mu\psi_{g}({\bf r}),
\label{ground_state_nlse}
\end{equation}
where $H_{0} = -\frac{\hbar^{2}}{2m}\nabla^{2} + V_{\rm trap}({\bf r})$ 
is the Hamiltonian for an isolated atom in the trap, $N_{0}$ is the
number of atoms in the condensate, $U_{0}$ represents the interaction 
between condensate atoms, and the eigenvalue $\mu$ is the chemical 
potential. 

In most current trap designs, $V_{\rm trap}$ can be described by the 
anisotropic harmonic oscillator potential, 
$V_{\rm trap}({\bf r}) = \frac{1}{2}m\left(\omega_{x}^{2}x^{2} + 
\omega_{y}^{2}y^{2} + \omega_{z}^{2}z^{2}\right)$, where $m$ is the 
atomic mass and $\omega_i = 2\pi \nu_i$ is the angular frequency of 
oscillation along the axis $i$.  The time--averaged orbiting potential 
(TOP) trap \cite{BEC!,TOP} treated here is cylindrically symmetric; 
its potential is given by $\omega_{x} = \omega_{y} = 
\omega_{\perp}$ and $\omega_{z} = \sqrt{8}\omega_{\perp}$.  The results
presented below correspond to $\nu_{z}=\omega_{z}/2\pi=210$ Hz.  The 
parameter $U_{0}$ expresses the interaction between two atoms as 
$U_{0} = 4\pi\hbar^{2}a/m$, where $a$ is the scattering length, which 
characterizes the zero-energy behavior of the $s$--wave phase shift in 
collisions between two atoms.  The scattering length $a$ is the only
piece of atomic collision data used as input to our calculations.  The 
present results are given in terms of the most recent~\cite{Heinzen} 
experimental value, $a=110 a_{0}$, where $a_{0}$ is the Bohr radius;
our calculations were actually carried out with a previously 
published~\cite{Rb87scatter} value of $a=100 a_{0}$, but since 
Eq.\ \ref{ground_state_nlse} obeys a scaling law involving $N_{0}$, 
$\nu_{\perp}$, and $a$ (see below), we can rescale our results to 
compare quantitatively with experiment.  It should be noted that for 
the alkali atoms in current BEC studies, the experimental determination
of $a$ requires extensive spectroscopic analysis, and present values 
are accompanied by substantial uncertainties~\cite{Tiesinga}.

Equation \ref{ground_state_nlse} has previously been solved by several
independent methods \cite{HC96,EDCRB96} to describe the BEC ground state.  
Here we investigate the response of the ground state to an oscillatory 
perturbation at angular frequency $\omega_p$.  The associated time-dependent 
GP equation takes the form:
\begin{equation}
i\hbar\frac{\partial\Psi}{\partial t} = 
\big[ H_{0} + U_{0}|\Psi({\bf r},t)|^{2}
+ f_{+}({\bf r})e^{-i\omega_{p} t} + 
f_{-}({\bf r})e^{i\omega_{p} t}\big]\Psi({\bf r},t),
\label{nlse}
\end{equation}
where $f_{\pm}({\bf r})$ are the spatially--dependent amplitudes of 
the perturbation.  We solve this equation in the linear--response limit.  
The details of this approach are described elsewhere \cite{iso_exc}, 
and we simply state the central results here.  By using the $Ansatz$
\begin{equation}
\Psi({\bf r},t) = e^{-i\mu t/\hbar}
\left[
N_{0}^{\frac{1}{2}}\psi_{g}({\bf r}) + 
u({\bf r})e^{-i\omega_{p} t} + v({\bf r})e^{i\omega_{p} t}\right].
\label{initial_form_of_wavefunction}
\end{equation}
we obtain the linear--response equations,
\begin{equation}
\left[{\cal L} - \hbar\omega_{p}\right]u({\bf r}) + 
N_{0}U_{0}\left[\psi_{g}({\bf r})\right]^{2}v({\bf r}) = 
-f_{+}({\bf r})\psi_{g}({\bf r}),
\label{u_equation}
\end{equation}
\begin{equation}
N_{0}U_{0}\left[\psi_{g}^{*}({\bf r})\right]^{2}u({\bf r}) +
\left[{\cal L} - \hbar\omega_{p})\right]v({\bf r}) = 
-f_{-}({\bf r})\psi_{g}({\bf r}),
\label{v_equation}
\end{equation}
where ${\cal L} = H_{0} - \mu + 
2U_{0}N_{0}\left|\psi_{g}({\bf r})\right|^{2}$.

This pair of equations can be solved by expansion in terms of the 
GP normal--mode equations,
\begin{equation}
\left[{\cal L} - \hbar\omega_{\lambda}\right]u_{\lambda}({\bf r}) + 
N_{0}U_{0}\left[\psi_{g}({\bf r})\right]^{2}v_{\lambda}({\bf r}) 
= 0,
\label{eigenmode_u_equation}
\end{equation}
\begin{equation}
N_{0}U_{0}\left[\psi_{g}^{*}({\bf r})\right]^{2}u_{\lambda}({\bf r}) +
\left[{\cal L} + \hbar\omega_{\lambda}\right]v_{\lambda}({\bf r})
= 0,
\label{eigenmode_v_equation}
\end{equation}
where $\omega_{\lambda}$ is an eigenvalue and $u_{\lambda}({\bf r})$, 
$v_{\lambda}({\bf r})$ are corresponding eigenfunctions.  For a 
homogeneous BEC ({\it i.e.} $V_{\rm trap}({\bf r}) = 0$), 
$\omega_{\lambda}$ would be a quasiparticle excitation frequency with
a continuous spectrum, corresponding to travelling sound waves.
In a trap, however, the elementary excitations of the BEC 
remained confined, so $u_{\lambda}({\bf r})$  and $v_{\lambda}({\bf r})$
must be square-integrable, and the spectrum of $\omega_{\lambda}$ 
is thus discrete.  The solution of eq. \ref{initial_form_of_wavefunction} 
can then be obtained in the ususal way by a superposition of 
eigenfunctions:
\begin{equation}
\left(
\begin{array}{c}
u({\bf r})\\
v({\bf r})
\end{array}
\right) = 
-\frac{1}{\hbar}\sum_{\lambda}\frac{g_{\lambda}}
{\omega_{\lambda} - \omega_{p}}
\left(
\begin{array}{c}
u_{\lambda}({\bf r})\\
v_{\lambda}({\bf r})
\end{array}
\right),
\label{linear_response_solutions}
\end{equation}
where
\begin{equation}
g_{\lambda} = \int d^{3}r\, 
\left[u_{\lambda}^{*}({\bf r})f_{+}({\bf r}) + 
v_{\lambda}^{*}({\bf r})f_{-}({\bf r})\right]\psi_{g}({\bf r}).
\label{mode_overlap}
\end{equation}
The response diverges when $\omega_{p}=\omega_{\lambda}$, {\it i.e.}
when the driving frequency is equal to a natural excitation frequency.  
In practice, of course, this divergence will be eliminated by damping 
effects, a subject of ongoing work that will not be discussed here.
To the extent that mechanisms of damping and their associated frequency 
shifts can be neglected, it is apparent that free oscillation of the 
condensate after a transient disturbance will occur at the frequencies 
$\omega_{\lambda}$.

As we have discussed elsewhere~\cite{iso_exc}, there is a straightforward 
connection between the resonant oscillation frequencies $\omega_{\lambda}$ 
and the the quasi--particle mode frequencies that are encountered in the 
standard Bogoliubov approximation~\cite{Fetter}.  Stated simply, 
Eqs.\ (\ref{eigenmode_u_equation}) and (\ref{eigenmode_v_equation}) are
identical to the equations that define the quasi--particle modes and
frequencies within the Bogoliubov approximation.  Thus an experiment that 
measures the free oscillatory response of a shaken BEC provides a direct 
observation of the quasi-particle spectrum.  In particular, by shaking the
BEC at a frequency near one of the resonances $\omega_{\lambda}$, one can 
produce a response that is dominated by the $\omega_{\lambda}$.  This is 
the approach that has been taken by the first such experiment 
reported~\cite{JILA}.

We have solved numerically the system of equations consisting of 
Eqs.\ (\ref{ground_state_nlse}), (\ref{eigenmode_u_equation}), and 
(\ref{eigenmode_v_equation}) under the conditions of that experiment.  
The solution was accomplished in two steps.  First, Eq.\ 
(\ref{ground_state_nlse}) was solved by expanding the solution 
$\psi_{g}({\bf r})$ in a basis set consisting of a finite number of 
trap eigenfunctions. The details of the numerical method have been 
recounted elsewhere~\cite{EDCRB96}.  Equations (\ref{eigenmode_u_equation}), 
and (\ref{eigenmode_v_equation}) were then solved by expanding 
$u_{\lambda}({\bf r})$  and $v_{\lambda}({\bf r})$ in the same basis set.  
These expansions convert Eqs.\ (\ref{eigenmode_u_equation}), and 
(\ref{eigenmode_v_equation}) into a generalized matrix eigenvalue problem 
that can be solved by standard numerical techniques~\cite{iso_exc}.  The 
error in the solutions was assessed by increasing the basis--set size 
until the mode frequencies converged to at least three figures.

\begin{figure}[htb]
\begin{center}
\mbox{\psboxto(3.1in;0in){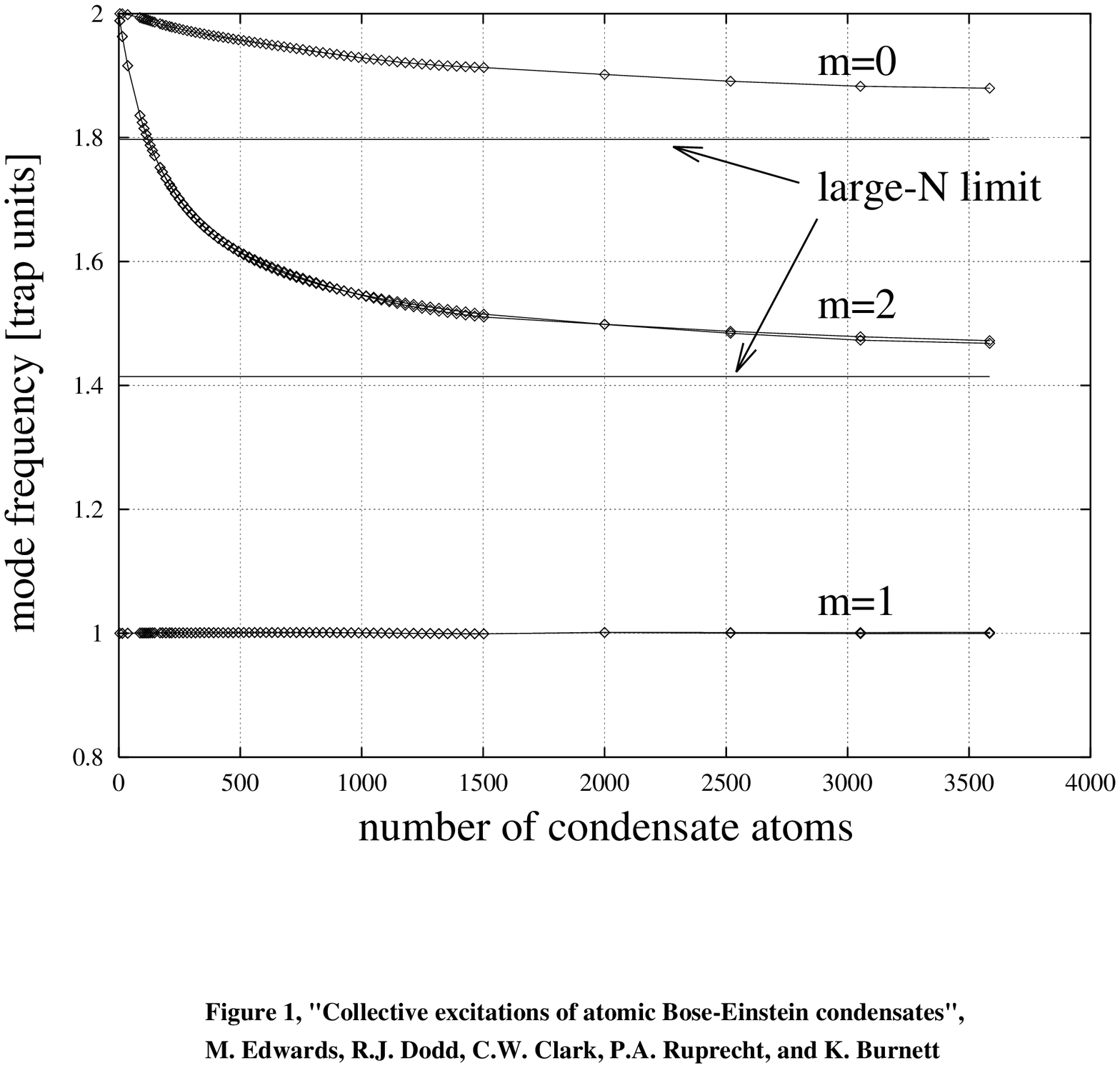}}
\end{center}
\caption{The lowest three calculated excitation
frequencies, in units of the perpendicular trap frequency
\protect{$\nu_{\perp}^{(t)}$}, of the JILA TOP--trap condensate,
{it vs.\ } the number of condensate atoms,  $N_{0}^{(t)}$.}
\label{spectrum}
\end{figure}

Figure~\ref{spectrum} shows our results for the lowest three excitation 
frequencies (in units of the trap frequency $\nu_{\perp}^{(t)}$), as a 
function of $N_{0}$.  A simple scaling law faciltiates comparison of 
calculation and experiment.  A solution of Eqs.\ (\ref{ground_state_nlse}), 
(\ref{eigenmode_u_equation}), and (\ref{eigenmode_v_equation}), plus 
normalization for experimental values of the parameters
$\{N_{0}^{(e)}, a_{e}, \nu_{\perp}^{(e)}, m_{e}\}$, will also 
satisfy the equations for the parameter set $\{N_{0}^{(t)}, a_{t}, 
\nu_{\perp}^{(t)}, m_{t}\}$, if the quantity
$\gamma = N_{0}a\left(m\nu_{\perp}\right)^{1/2}$
is constant.  Thus, excitation measurements performed on a BEC 
with $N_{0}^{(e)}$ atoms in a trap of frequency $\nu_{\perp}^{(e)}$ 
can be related to the spectrum displayed in Fig. \ref{spectrum}
by taking
\begin{equation}
N_{0}^{(t)} = 
\left(
\frac{a_{e}}{a_{t}}
\right)
\left(
\frac{\nu_{\perp}^{(e)}}{\nu_{\perp}^{(t)}}
\right)^{1/2}
N_{0}^{(e)},
\label{new_prediction}
\end{equation}
where $\nu_{\perp}^{(t)}=210/\sqrt{8}\approx 74.25$ Hz. 

We have used this scaling law to compare our results with those of the
recent experiment \cite{JILA} where excitations of an atomic BEC were 
observed for the first time.  Table \ref{table1} presents the comparison. 
No attempt has been made here to account for experimental uncertainties. 
The ranges presented for the experimental ratio 
$\left(\nu_{mode}^{(e)}/\nu_{\perp}^{(e)}\right)$ involve a 1--3 \% 
reduction of the raw data to extrapolate to zero--amplitude driving 
force~\cite{eac_pc}.  It is fair to say that the agreement is excellent, 
as the difference between theory and experiment ranges from 2--6 \%.

To better understand the nature of these excitations we compare them with
the results of Stringari~\cite{Stringari} who obtained analytic solutions
to the linearized GP equation in the hydrodynamic limit ($N_{0}\rightarrow
\infty$).  The curve labeled ``$m=1$'' is a doubly--degenerate dipole 
excitation that coincides exactly with the first excited state of the bare 
trap.  This is because the lowest dipole mode of an ensemble of identical 
interacting atoms in an external harmonic potential corresponds to a rigid 
motion of the center--of--mass, independent of the nature of interatomic 
forces~\cite{Stringari}.  

\begin{figure}[htb]
\begin{center}
\mbox{\psboxto(3.2in;0in){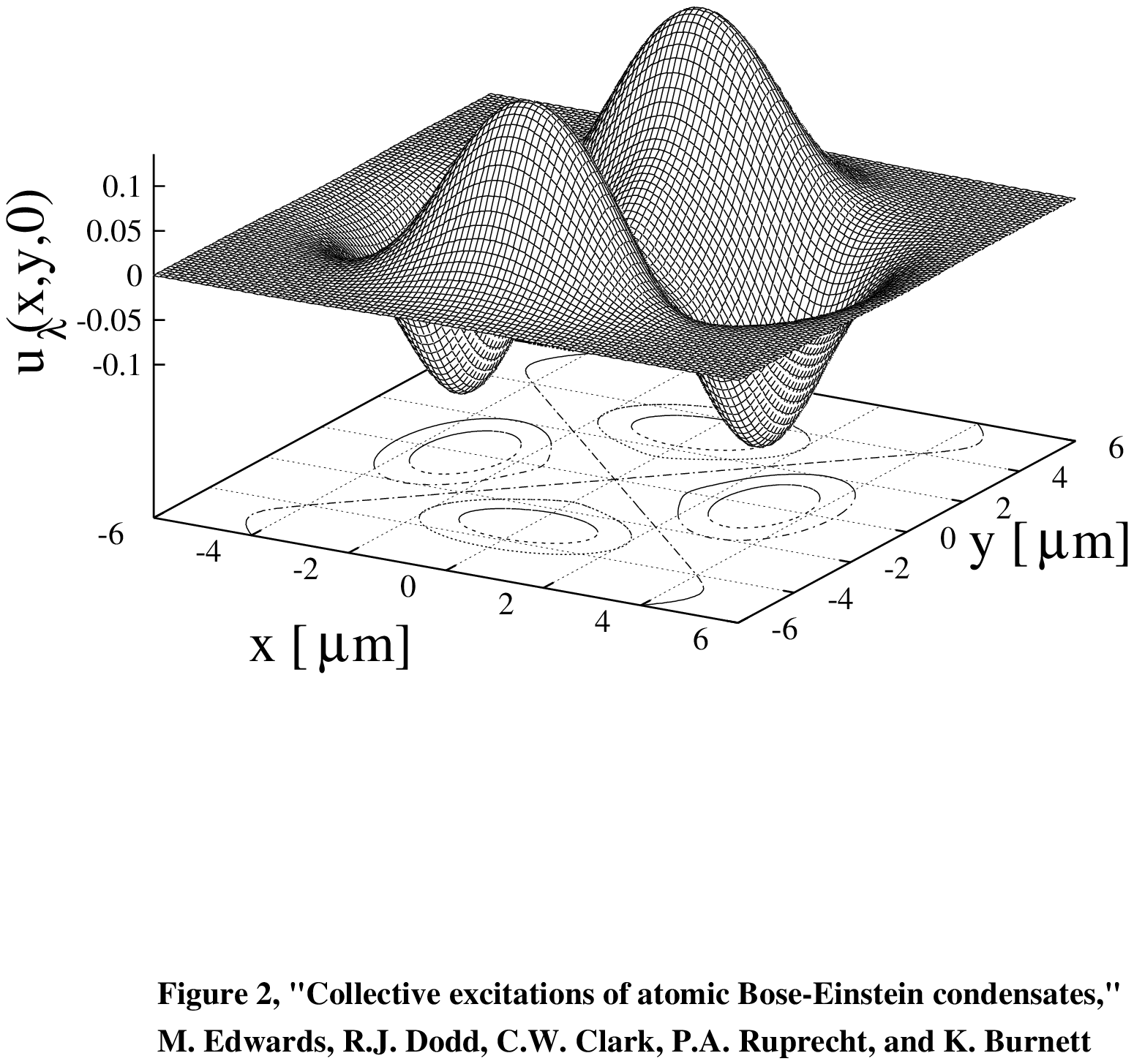}}
\end{center}
\caption{The $u_{\lambda}$--component of the $m=|2|$ excitation 
(middle curve of fig.\ \protect{\ref{spectrum}}), displaying its
quadrupolar shape.}
\label{mode}
\end{figure}

The curve labeled ``$m=2$'' corresponds to doubly degenerate excitations 
that correlate to the lowest $m=|2|$ energies of the two-dimensional 
harmonic oscillator in the noninteracting limit.  The quadrupolar nature 
of these excitations is exhibited in fig.\ \ref{mode}, which contains a 
plot of $u_{\lambda}$ over a region of the $xy$ plane [{\it i.e.\ } 
$u_{\lambda}(x,y,0)$] for $N_{0}=2000$ atoms.  This plot clearly shows the 
four--peak structure characteristic of quadrupole excitations.  The 
large--$N_{0}$ limit of this mode, 
$\nu_{mode}/\nu_{\perp}\rightarrow\sqrt{2}$, is also shown in figure 
\ref{spectrum}.  The fact that the middle curves also seem to be approaching 
this limit as $N_{0}\rightarrow\infty$ also lends support to their 
interpretation as quadrupole excitations.  

The curve labeled ``m=0'' corresponds to a ``large--$N_{0}$'' excitation 
that is a breathing mode in the $xy$ plane.  The asymptotic limit is 
shown in fig.\ \ref{spectrum} and is given by~\cite{Stringari}
\begin{equation}
\nu_{mode}/\nu_{\perp}\rightarrow 
\left(
2 + \frac{3}{2}\lambda^{2} - 
\frac{1}{2}\sqrt{9\lambda^{4} - 16\lambda^{2} + 16}
\right)^{1/2}
\label{mono_+_quad}
\end{equation}
where $\lambda=\nu_{z}/\nu_{\perp}=\sqrt{8}$ for the TOP trap.  As
discussed in ref.\cite{JILA}, the symmetries of the normal modes can
be tested by experimental selection rules, and the classifications
of the observed modes are found to agreement with those given here.

In conclusion, we have presented excitation spectra that agree well with
the data of a recent experiment.  We have shown that these data constitute 
a direct measurement of the $T=0$ Bogoliubov spectrum of an atomic BEC.  We 
have also used the large--$N_{0}$ limit and mode shapes to describe the 
nature of these excitations.  The experimental confirmation of these data 
will have significant implications for understanding of the many--body 
physics of these dilute, weakly--interacting bosonic systems, and for 
practical use in future BEC engineering.

\acknowledgments 

We thank D.\ S.\ Jin, J.\ R.\ Ensher, M.\ R.\ Matthews, C.\ E.\  Wieman, 
and E.\ A.\ Cornell for stimulating discussions and for making their 
experimental data available to us in advance of publication.  Work at 
Oxford was supported with funding from the Rhodes Trust and the 
U.\ K.\ Engineering and Physical Sciences Research Council.  M.\ Edwards 
acknowledges funding from National Science Foundation grant PHY-9505468 
and support from the GSU Foundation.  This work was supported in part by 
the Institute for Theoretical Atomic and Molecular Physics at Harvard 
University and the Smithsonian Astrophysical Observatory.

\newpage

\newpage

\begin{table}
\caption{This table presents the comparison with excitation data obtained
in the experiment of ref.\ [4].}
\label{table1}
\begin{tabular}{cccccccc}
mode&$\nu_{\perp}^{(e)}$&$\nu_{mode}^{(e)}$&$N_{0}^{(e)}$&
$\frac{\nu_{mode}^{(e)}}{\nu_{\perp}^{(e)}}$&
$N_{0}^{(t)}$&$\frac{\nu_{mode}^{(t)}}{\nu_{\perp}^{(t)}}$
&\% diff.\\
\tableline
$m=0$&43.2&82&3420&1.84---1.88&2870&1.89&2.7\\
$m=0$&132&244.7&2400&1.79---1.83&3520&1.88&5.0\\
$m=2$&43.2&62.5&2800&1.41---1.44&2350&1.49&5.7\\
$m=2$&132&188.5&2200&1.39---1.42&3230&1.47&5.8\\
\end{tabular}
\end{table}

\end{document}